\documentclass[PPCF]{iopart}
\usepackage{graphicx}
\usepackage{amssymb}
\usepackage{breqn}

\begin{document}

\title{Global constant field approximation for radiation reaction in collision of high-intensity laser pulse with electron beam}

\author{I.~I.~Artemenko}
\author{M.~S.~Krygin}
\author{D.~A.~Serebryakov}
\author{E.~N.~Nerush}
\ead{nerush@appl.sci-nnov.ru}
\author{I.~Yu.~Kostyukov}
\address{Institute of Applied Physics of the Russian Academy of
Sciences, 46 Ulyanov St., Nizhny Novgorod 603950, Russia}

\begin{abstract}
    In the laser --- electron beam head-on interaction electron energy can decrease due to
    radiation reaction, i.e. emission of photons. For 10--100~fs laser pulses and for the laser
    field strength up to the pair photoproduction threshold, it is shown that one can calculate the
    resulting electron and photon spectra as if the electron beam travels through a constant
    magnetic field. The strength of this constant field and the interaction time are found as
    functions of the laser field amplitude and duration. Using of constant field approximation can
    make a theoretical analysis of stochasticity of the radiation reaction much more simple in
    comparison with the case of alternating laser field, also, it allows one to get electron and
    photon spectra much cheaper numerically than by particle-in-cell simulations.
\end{abstract} \ioptwocol

\maketitle

\section{Introduction}
\label{Intro}

The collision of light from optical and free-electron lasers with ultrarelativistic electron beams is being used
routinely nowadays to produce MeV photons via Compton scattering on facilities such as
HI$\gamma$S~\cite{Weller09, HIGS} and NewSUBARU~\cite{Ando98, NewSUBARU}. This method is also a
central point of ELI-NP facility~\cite{Adriani14, ELINP} which is planned to become operational as user
facility in 2019~\cite{Balabanski18}.

The theory of scattering of light by relativistic electrons is
well known, however, in some interaction regimes only a few experiments are carried out,
e.g. in nonlinear and quantum regimes.
At high intensity of the laser radiation
an electron absorbs a big number of optical photons in order to produce a single hard
photon --- the scattering becomes nonlinear~\cite{Sarri14, Yan17}.
At even higher intensities of laser radiation the photon recoil occurs, and the photon emission
obeys quantum synchrotron formulas~\cite{Berestetskii82, Elkina11}, testing of which is a
long-standing experimental challenge~\cite{Burke97, Bamber99, Cole18, Poder18}. The interest to
such studies was fueled recently by study~\cite{Wistisen18} of quantum radiation reaction in aligned crystals that demonstrates
a discrepancy between experimental data and results obtained with the quantum synchrotron formula.
Thus, a discussion of theoretical models of radiation reaction reopens~\cite{Macchi18, Piazza19}.

In the context of intense laser pulses with $a_0 \gg 1$, 
the photon emission by relativistic electrons and the corresponding modification of the electron
spectrum (i.e. radiation reaction) is generally treated by particle-in-cell (PIC)
simulations with
Monte Carlo (MC) technique incorporated~\cite{Elkina11, Nerush11b, Ridgers14, Vranic16, Niel18a} (here $a_0 = e E_0 / m c
\omega_L$ with $E_0$ the electric field amplitude, $e > 0$ the elementary charge, $m$ electron mass
and $c$ the speed of light). PIC-MC simulations solves implicitly the Boltzmann's
equations (BEs) for particle distribution functions~\cite{Elkina11, Nerush11c} which can
be solved directly in numerical simulations as well~\cite{Bulanov13a}. 

It is shown in Ref.~\cite{Bulanov13a} that if the electron energy remains high enough during the
collision, the electrons trajectories are bended negligibly and can be considered straight. For the
sake of further simplification one can assume that the transverse size of the electron beam is
smaller than the laser spot size, and also neglect the collective (plasma) effects in the electron beam.
In this case the radiation reaction is described within
one-dimensional (1D) BE:
\begin{equation}
    \label{Boltzmann}
    \partial_t f_e = -U f_e + \int_\epsilon^\infty f_e(\epsilon') w(\epsilon' \to \epsilon) \, d\epsilon',
\end{equation}
where, unless otherwise specified, all quantities are given at $(t, \epsilon)$, also,
$w(\epsilon' \to \epsilon) \, d\epsilon \, dt$ is the probability for an electron with energy
$\epsilon'$ in a time interval $dt$ to emit a
photon such that the resulting electron energy is in the interval $(\epsilon, \epsilon +
d\epsilon)$, with $d\epsilon$ and $dt$ infinitesimals. Note that it is possible to describe the
evolution of the electron distribution with~\eref{Boltzmann} because the emission probability $w$
is determined by the local field value~\cite{Berestetskii82} and no interference of the emitted photons occurs.
Here $U$ is the full emission probability rate
\begin{equation}
    U(\epsilon) = \int_0^\epsilon w(\epsilon \to \epsilon') \, d\epsilon'.
\end{equation}

The photon distribution function $f_{ph}$ do not take part in~\eref{Boltzmann} in the classical
radiation reaction limit, i.e. until the
quantum parameter
\begin{equation}
    \chi = \epsilon b
\end{equation}
is small, $\chi \lesssim 1$, hence no pair photoproduction occurs~\cite{Elkina11}. Here $b = F / e E_S$ is the ratio
of the Lorentz force transverse to the electron velocity [$\mathbf F \approx | \mathbf{ F_L \times v / c}
|$ with $\mathbf F_L = e ( \mathbf{ E + v \times B} / c)$] to the force in the Sauter--Schwinger field $E_S = m^2 c^3 /
e \hbar$ with $\mathbf v$ the electron velocity, $\mathbf E$ and $\mathbf B$ the electric
and magnetic fields, respectively,
$\hbar = h / 2 \pi$ and $h$ the Planck's constant. From here on the electron energy $\epsilon$
is normalized to the electron rest energy $m c^2$.

The photon emission probability depend on the laser field, which varies with time along the
electron trajectory,
and the general solution of~\eref{Boltzmann} reads as
\begin{equation}
    \label{solution}
    f_e(t) = \mathrm{T}\exp \left[\int_0^t \hat A(t') \, dt' \right] \, f_e(0),
\end{equation}
where the linear operator $\hat A$ represents the right hand side of~\eref{Boltzmann}
(multiplication by $U$ and convolution with $w$), and $\mathrm{T} \exp$ is the time-ordered
exponential known, for example, from the quantum field theory~\cite{Berestetskii82}. It can be
considered as a Taylor series of the exponential function, with ordered products of $\hat A(t_1) \hat A(t_2)$
such that $t_2 > t_1$. The time-ordered exponential can be found
numerically~\cite{Bulanov13a}, whereas its theoretical analysis in the general case faces the
difficulties similar to that in Lie theory (see, for instance, Campbell--Hausdorff
formula~\cite{Kirillov08}).

If the laser pulse is not single-cycled, the rising and descent slopes of a half-wave have the same
shape and they yield in~\eref{solution} sum of products of the same operators in the opposite order, $\hat
A(t_1) \hat A(t_2) + \hat A(t_3) \hat A(t_4)$ with $\hat A(t_3) = \hat A(t_2)$ and $\hat A(t_4)
= \hat A(t_1)$. Thus, one can assume that the commutator of $\hat A(t_1)$ and $\hat A(t_2)$ is averaged away for any $t_1$ and
$t_2$. In this case $\mathrm{T} \exp$ is the ordinary
exponential and the integration in~\eref{solution} can be considered as a sort of averaging
(that is especially clear for the matrix representation, see below).
Therefore, in this case the general solution of BEs (with alternating field) given at $t$ is
equivalent to the solution of BEs with some new functions $U^*$ and $w^*$
which do not depend on time on the interval $[0, t]$.

One can try to use $w$ and $U$ for some global constant magnetic field and some interaction time
instead of the averaged probabilities of the emission in the laser field $w^*$ and $U^*$. The
correspondence between the alternating laser field and the constant magnetic field which result the same
electron
distribution function (from here on global constant field approximation, GCFA) can be justified
rigorously in the Fokker--Planck (FP) regime of the radiation reaction. Indeed, 
in the FP regime~\cite{Niel18a, Niel18b}, when the energy of individual photons is small
and the number of the emitted photons per electron is large, the resulting electron spectrum do not
depend on the exact shape of $w(\epsilon' \to \epsilon)$, and depend only on the first and
second momenta of $w$. Thus, one can easily find the parameters of GCFA (the value of the constant field, and the
interaction time) which lead to the same distribution function $f_e(t)$ as for the
laser field, as shown in \Sref{FP}. Then in \Sref{Scintillans} the matrix representation of BEs
and the numerical method to solve them for a constant magnetic field, are considered. In
\Sref{Num} GCFA is tested in FP regime as well as in the regime of
low number of the emitted photons and in the quantum regime $\chi \sim 1$. Moreover, here we
compare the photon spectrum obtained in the laser field with that in the constant magnetic field.

\section{GCFA in Fokker--Planck regime of radiation reaction}
\label{FP}

It is shown by Baier and Katkov~\cite{Baier67, Baier98, Berestetskii82} that the
photon emission in the quantum synchrotron regime in almost arbitrary field is described by the same
formulae as for a pure magnetic field,
\begin{eqnarray}
    \label{w}
    w(\epsilon \to \epsilon') = -\frac{\alpha m c^2}{\hbar \epsilon^2} \left[ \int_z^\infty
    \mathrm{Ai}(\xi) \, d \xi + \vartheta \mathrm{Ai}'(z) \right], \\
    \vartheta = \frac{2}{z} + \frac{(\epsilon - \epsilon') \chi
    z^{1/2}}{\epsilon}, \\
    z = \left( \frac{\epsilon - \epsilon'}{\epsilon' \chi} \right)^{2/3},
\end{eqnarray}
with $\alpha = e^2 / \hbar c \approx 1 / 137$ the fine-structure constant.
If the quantum parameter $\chi \ll 1$ (so-called classical limit), it is seen from~\eref{w}
that the emission spectra is determined by the critical frequency $\omega_c$,
\begin{equation}
    \frac{\hbar \omega_c}{mc^2} = \chi \epsilon,
\end{equation}
namely, the width of the emission spectra and the mean energy of the emitted photons is of the order
of $\hbar \omega_c$~\cite{LandauII}. Furthermore, the full emission probability is $U \sim \alpha / t_{rf}$ with
\begin{equation}
    t_{rf} \sim m c / F
\end{equation}
is the radiation formation time. Thus, in the average
an electron emits a single photon in a time of about $137 \, t_{rf}$. Note that in
the classical limit $\hbar \omega_c \ll mc^2 \epsilon$.

One can turn from the differential equation~\eref{Boltzmann} to the consideration
of the photon emission as a stochastic process~\cite{Bashinov15}.
The electron energy decreases every time when the electron emits a photon. Thus, the final electron energy
is its initial energy minus the energies of the photons that the electron have emitted.
In the FP regime, as long as the number of the emitted photons is large and the photon energy is much smaller
than the electron energy, the different events of the photon emission can be considered
independently, and the central limit theorem can be applied. Hence the mean drop of the electron
energy and the electron energy variance ($\mu$ and $\sigma^2$, respectively) are the sums of these quantities for
individual photon emission events.
The critical frequency is proportional to $F$, hence for the emission of a single photon its mean
energy $\mu_i \propto F$ and the energy variance is $\sigma_i^2 \propto F$. Recalling that the full
emission probability $U \propto \omega_c \propto F$, and replacing the sum in the central limit
theorem with the integral, one get
\begin{eqnarray}
    \label{mu}
    \mu \propto \int_0^t F^2(t') \, dt', \\
    \label{sigma}
    \sigma^2 \propto \int_0^t F^3(t') \, dt'.
\end{eqnarray}
Equations~\eref{mu} and~\eref{sigma} allow one readily find the correspondence between the
constant magnetic field and the laser field such that $\mu$ and $\sigma$ are the same in both cases. Note that in
the above-mentioned considerations we implicitly suppose a fixed number of the emitted photons in
the central limit theorem.

To compute GCFA parameters definitely, one can consider two setups. The first is a passing of an electron through
the magnetic field $\mathcal H_0$ such that $F = e \mathcal H_0$ during the time interval $\tau$. The second is the head
on collision of an electron
with a finite plane wave in that the electric field is
\begin{equation}
    \label{Elaser}
    E = E_0 \cos^2 \left( \frac{\pi \xi }{2 x_L} \right) \cos(k_L \xi),
\end{equation}
and the magnetic field is $H = E$, with $\xi = x - ct$, $k_L = \omega_L / c$, $\omega_L$ the circular
frequency and $x_L$ the FWHM of the electric field.
If the initial
electron position is $x(0) = x_L$, then the full passing of the electron through the pulse
corresponds to $t \in [0, x_L / c]$ and
\begin{equation}
    F \approx 2 e E_0 \sin^2 \left( \frac{\pi c t}{x_L} \right) \cos(k_L x_L - 2 \omega_L t).
\end{equation}

Considering the same initial energy of the electrons and demanding the same $\mu$ and $\sigma$ in both
cases, from~\eref{mu} and \eref{sigma} one get:
\begin{eqnarray}
    \label{corr1}
    \mathcal H_0  = \frac{40}{9 \pi} E_0, \\
    \label{corr2}
    \tau = \frac{243 \pi^2}{6400} x_L / c,
\end{eqnarray}
where we assume $x_L \gg 2 \pi / k_L$.

\section{Numerical integration of Boltzmann's equations in constant magnetic field}
\label{Scintillans}

For numerical solution of BE in a constant magnetic field the code \textit{Scintillans} is
presented~\cite{Scintillans}
which is available under the BSD3 licence. The code is written in Haskell language and is based on
the \textsc{repa} library~\cite{Lippmeier13} that results a C-comparable performance. However, the
\textit{Scintillans} is written to be not the most performant but simple and straightforward first. The version 0.3.0
used here solves BEs
which includes electrons and photons~\cite{Elkina11, Bulanov13a},
\begin{eqnarray}
    \label{BE1}
    \partial_t f_e = -U f_e + \int_\epsilon^\infty f_e(\epsilon') w(\epsilon' \to \epsilon) \,
    d\epsilon', \\
    \label{BE2}
    \partial_t f_{ph} = \int_\epsilon^\infty f_e(\epsilon') w(\epsilon' \to \epsilon' - \epsilon) \, d\epsilon'.
\end{eqnarray}
For this the following numerical representation is used:
\begin{eqnarray}
    \partial_t f = \hat A f, \\
    f = \left(
    \begin{array}{c}
        f_e \\
        f_{ph}
    \end{array} \right), \\
    \hat A =
    \left( \begin{array}{cc}
        \hat A_{00} & \hat A_{01} \\
        \hat A_{10} & \hat A_{11}
    \end{array} \right),
\end{eqnarray}
where $f_e$ and $f_{ph}$ are the column vectors representing the electron and photon
distribution functions on the energy intervals $[\epsilon_a, \epsilon_b]$ and $[0, \epsilon_b -
\epsilon_a]$:
\begin{eqnarray}
    f_e = \left( \begin{array}{c}
               f_e(\epsilon_a)                   \\
               f_e(\epsilon_a + \Delta \epsilon) \\
               \vdots                            \\
               f_e(\epsilon_b)
    \end{array} \right) \\
    f_{ph} = \left( \begin{array}{c}
               f_{ph}(0)                    \\
               f_{ph}(\Delta \epsilon)      \\
               \vdots                       \\
               f_{ph}(\epsilon_b - \epsilon_a)
    \end{array} \right)
\end{eqnarray}
which contain $n$ nodes each. Here $\Delta \epsilon = (\epsilon_b - \epsilon_a) / (n - 1)$.
Using different but consistent energy intervals for the electron and photon distribution
functions is especially useful in the classical regime where one can choose $\epsilon_b -
\epsilon_a \ll \epsilon_b$.

The elements of the block matrix $\hat A$ are the matrices $\hat A_{00} = -\hat U + \hat W$, $\hat
A_{01} = \hat A_{11} = \hat 0$ (zero matrix) and $\hat A_{10}$. The
matrices $\hat U$, $\hat W$ and $\hat A_{10}$ are defined as follows. First,
\begin{equation}
    \hat W = \left( \begin{array}{ccccc}
        w_{0 \to 0} & w_{1 \to 0} & w_{2 \to 0} & w_{3 \to 0} & \cdots \\
        0           & w_{1 \to 1} & w_{2 \to 1} & w_{3 \to 1} & \cdots \\
        0           & 0           & w_{2 \to 2} & w_{3 \to 2} & \cdots \\
        0           & 0           & 0           & w_{3 \to 3} & \cdots \\
        \cdots      & \cdots      & \cdots      & \cdots      & \cdots
    \end{array} \right), \\
\end{equation}
where $w_{k \to l} = w(\epsilon_a + k \Delta \epsilon \to \epsilon_a + l \Delta \epsilon)$. Note that the
transitions with negative indices (e.g. $w_{0 \to -1}$) are not taken into account. Thus, it is
assumed that $\epsilon_a$ is such small that the photon emission is negligible for electrons with
$\epsilon_a$, or there is almost
no electrons with energy about $\epsilon_a$. Then, the matrix $\hat U$ is the diagonal matrix with
the diagonal elements computed as sums within the corresponding columns of $\hat W$. To compute $\hat A_{10}$ we
permute the elements in the columns of $\hat W$:
\begin{equation}
    \hat A_{10} = \left( \begin{array}{ccccc}
        w_{0 \to 0} & w_{1 \to 1} & w_{2 \to 2} & w_{3 \to 3} & \cdots \\
        0           & w_{1 \to 0} & w_{2 \to 1} & w_{3 \to 2} & \cdots \\
        0           & 0           & w_{2 \to 0} & w_{3 \to 1} & \cdots \\
        0           & 0           & 0           & w_{3 \to 0} & \cdots \\
        \cdots      & \cdots      & \cdots      & \cdots      & \cdots
    \end{array} \right).
\end{equation}

The singular behavior of $w(\epsilon \to \epsilon')$ (it tends to infinity if $\epsilon'$ tends to
$\epsilon$) is not important here because the evolution of $f_e$ is determined rather by the
emission power distribution $(\epsilon - \epsilon') \, w(\epsilon \to \epsilon')$ which is not
singular. Thus, in the numerical approximation it is enough to resolve scales about some fraction
of the
critical frequency (in the classical limit) or about some fraction of $\epsilon$ (in the quantum
limit). For the
same reason we use $w_{k \to k} = 0$.

Equations~\eref{BE1} and~\eref{BE2} now can be solved with Euler's method
\begin{eqnarray}
    \label{solSc}
    f(t) = \exp(\hat A t) f(0), \\
    \label{expSc}
    \exp(\hat A t) \approx (\hat 1 + \hat A \Delta t)^p,
\end{eqnarray}
with $p = [t / \Delta t]$ and $\hat 1$ the identity matrix. The exponentiation in~\eref{expSc} can be performed
with exponentiation-by-squaring algorithm which has $O(\log p)$ complexity. Hence, the
numerical solution of BEs in GCFA
is much faster than the computation of BE solution for a variable field [with comlexity $O(p)$].
It is also obvious that the
solution of 1D BEs in GCFA is much faster than PIC computations.

Finally, \eref{solSc} and~\eref{expSc}, as the original BEs~\eref{BE1} and~\eref{BE2}, conserve the
number of the electrons and
the overall energy. Namely, the multiplication of $(\hat 1 + \hat A_{00} \Delta
t)$ and $f_e$ do not modify the number of the electrons,
\begin{dmath*}
    {\sum_{k, l} (\hat 1 + \hat A_{00})_{kl} (f_e)_l = \sum_{k, l} [\hat 1 + (\hat W - \hat U)
    \Delta t]_{kl}
    (f_e)_l}
    = {\sum_l (f_e)_l \equiv N_e},
\end{dmath*}
because $\hat U$ is computed from $\hat W$ and the sum along any column of $\hat A_{00}$ is zero.
Then, the change of the overall energy caused by the multiplication by $(\hat 1 + \hat A \Delta t)$
is zero,
\begin{equation}
    \sum_{k,l} \left[ (\epsilon_a + k \Delta \epsilon) (A_{00})_{kl} + k \Delta \epsilon
    (A_{10})_{kl} \right] (f_e)_l \Delta t = 0,
\end{equation}
because the summation by $k$ yields
\begin{dmath*}
    \sum_{k = 0}^{n - 1} k \left[ (A_{00})_{kl} + (A_{10})_{kl} \right] = 
    {-U_{ll} l + \sum_{k = 0}^l ( w_{l \to k} + w_{l \to (l - k)} ) k} = \left( \sum_{k = 0}^l w_{l
    \to (l - k)} - U_{ll} \right) l + {\sum_{k = 0}^l \left[k w_{l \to k} - (l - k) w_{l \to (l -
    k)} \right] = 0}.
\end{dmath*}
Therefore, despite of its simplicity, \textit{Scintillans} provides fast and robust numerical
solver of
BEs for the electrons and photons in a constant
magnetic field.

\section{GCFA in Fokker--Planck regime and beyond: comparison with PIC simulations}
\label{Num}

\begin{figure}
	\includegraphics[width=1\linewidth]{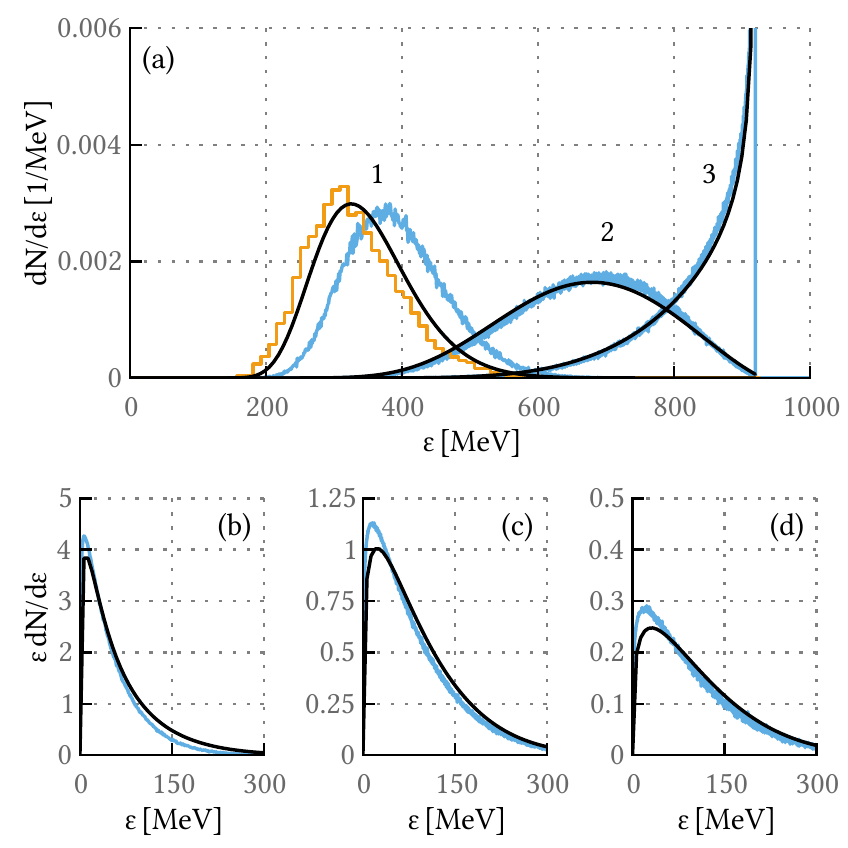}
    \caption{\label{fokkerplanck} Particle spectra computed with PIC-MC technique (light noisy lines) in
    the laser pulse --- electron beam collision for various pulse duration, and spectra computed in
    GCFA (black smooth lines), i.e. from Boltzmann's equations in a constant magnetic field. The
    magnetic field strength and the interaction time relates with the laser field parameters
    according to~\eref{corr1}-\eref{corr2}, yielding $\chi = 0.1$ in GCFA. Lines 1, 2 and 3
    demonstrate the electron spectra, and figures (b), (c) and (d) demonstrate the photon energy
    spectra for the laser pulse duration of $270$, $68$ and $17 \; \mathrm{fs}$, respectively.
    The orange stepped line 1 in figure (a) shows the result of PIC simulation for a laser pulse with no
    diffraction. See text for details.}
\end{figure}

To test GCFA, three dimensional (3D) simulations of the electron beam collision with the laser
pulse are performed with PIC code \textsc{quill}~\cite{QUILL}. The code \textsc{quill} takes into account photon
emission by the Monte Carlo
approach similar to rejection sampling~\cite{Elkina11, Nerush17}. In the simulations the laser pulse longitudinal shape is given
by~\eref{Elaser}, and
additional transverse envelope introduced which FWHM along the $y$ and $z$ axes are $y_s$
and $z_s$, respectively. The duration $x_L / c$ and the amplitude of the laser pulse, as well as
the initial electron energy $\epsilon_0$, are varied to obtain
different regimes of the radiation reaction. The interaction time allows complete propagation of
all the electrons through the laser pulse. The full electron beam radius ($2 \lambda_L$) is much smaller than the
laser pulse transverse size ($y_s = z_s = 13.4 \lambda_L$), where $\lambda_L = 2 \pi c / \omega_L =
1 \;
\mu \mathrm{m}$ is the laser wavelength.

The results of PIC simulations for unnormalized initial electron energy $\varepsilon_0 =
920 \; \mathrm{MeV}$ ($\epsilon = 1800$) and $a_0 = 16.5$ are shown in \fref{fokkerplanck} with
light lines. The maximal quantum parameter
reached in this case is $\chi \approx 0.07$. For $x_L = 82 \lambda_L \gg 137 ct_{rf}$ ($x_L / c =
270 \; \mathrm{fs}$, $137 t_{rf} \approx 4.4 \; \mathrm{fs}$) FP regime is attained. Accordingly to
the central limit theorem, in this case the electron
spectrum is close to the Gaussian one [noisy blue line $1$ in \fref{fokkerplanck}(a)]. The corresponding photon
energy spectrum $\epsilon f_{ph}(\epsilon)$ is shown as a light blue line in \fref{fokkerplanck}(b).
Similarly, the electron spectra for
$4$ and $16$ times shorter laser pulses ($x_L / c = 68$ and $17 \; \mathrm{fs}$, respectively) are
shown as light blue lines $2$ and $3$ in \fref{fokkerplanck}(a), and
the corresponding photon energy spectra are shown in figures~\ref{fokkerplanck}(c) and (d). Note that for the short
laser pulse every electron emits a few photons in the average that makes FP
approach~\cite{Niel18a}
and the central limit theorem inapplicable.

The results obtained
with \textit{Scintillans} in GCFA are shown in \fref{fokkerplanck} as black lines.
Equations~\eref{corr1} and~\eref{corr2}
are used to obtain the strength of the constant magnetic field and the interaction time from the
parameters of the laser pulses. Thus, black lines $1$, $2$ and $3$ in \fref{fokkerplanck}(a) are the
electron spectra, and in figures~\ref{fokkerplanck}(b), (c) and (d) are the photon energy
distribution $\epsilon f_{ph}(\epsilon)$ for $\mathcal H_0 = 250 \; \mathrm{kT}$ and $\tau = 102$,
$26$ and $6.4 \; \mathrm{fs}$, respectively. These parameters yield $\mathcal H_0 / E_S
= 5.6 \times 10^{-5}$, the initial value of $\chi = 0.1$ and the photon emission time $137 t_{rf} = 3.1 \;
\mathrm{fs}$.

In \textit{Scintillans}
simulations, as well as in PIC simulation, the electron distribution function is normalised in the
same way
\begin{equation}
    \int_0^\infty f_e(\epsilon) \, d\epsilon = 1.
\end{equation}
Note that the parameters of the \textit{Scintillans} simulation for dark curve $1$ in \fref{fokkerplanck}(a) are
the same as for middle curves in figure~10(c) from Ref.~\cite{Niel18a}, except here initially
monoenergetic electron beam with zero temperature is used. Nevertheless, the curve in
\fref{fokkerplanck}(a) is quite close to the corresponding curves from Ref.~\cite{Niel18a}
indicating the consistency between \textit{Scintillans} and \textsc{smilei} simulations.

For the longest laser pulse an additional PIC simulation is performed in which the laser field is
defined analytically in the form~\eref{Elaser}, with no transverse envelope hence no diffraction.
The resulting electron spectrum [orange stepped line 1 in \fref{fokkerplanck}(a)] is close to that
obtained in GCFA, thus the discrepancy between blue and black lines 1 in \fref{fokkerplanck}(a) is
caused by the diffraction. Therefore, it can be concluded from \fref{fokkerplanck} that the
electron and photon spectra can be described well in GCFA in the classical limit if 1D BE is
applicable, even if the number of the emitted photons is small.

\begin{figure}
	\includegraphics[width=1\linewidth]{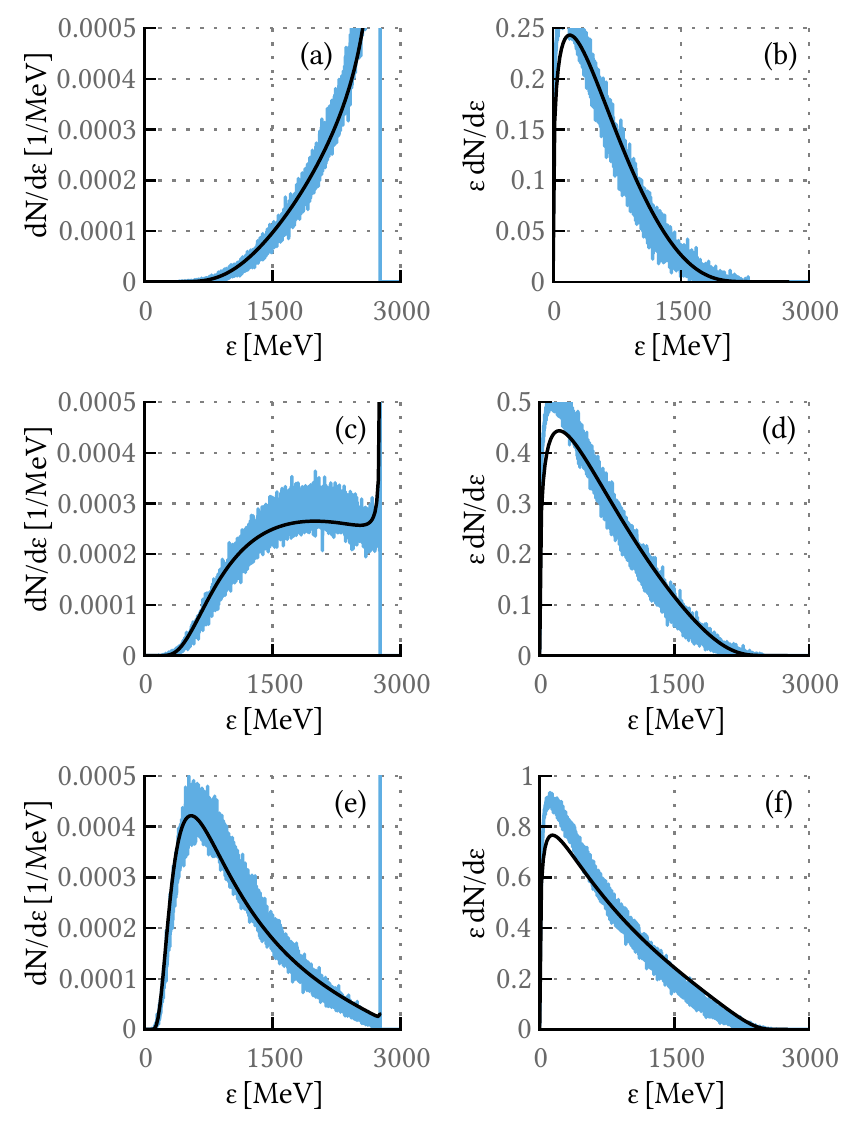}
    \caption{\label{higherchi} The electron spectrum (left column) and the photon energy spectrum
    (right column) in laser pulse --- electron beam collision computed with PIC-MC technique (light
    blue
    lines) and in GCFA (black smooth lines). The laser pulse amplitude varies such that in GCFA
    initially $\chi = 0.3$ for figures (a, b), $\chi = 0.6$ for figures (c, d) and $\chi = 1.2$ for
    figures (e, f). See text for details.}
\end{figure}

Fairly good coincidence between the electron and photon spectra computed in the constant magnetic field and in
the laser field forces to test GCFA in the regime of moderate $\chi$ values, $\chi \sim 1$.
For this, one should use quite short laser pulses, $x_L \lesssim 137 t_{rf}$, otherwise the electrons
loose almost all initial energy making 1D BE inapplicable. Note that at $\chi \gtrsim 1$ pair
photoproduction (which is not taken into account in \textit{Scintillans} simulations) and
subsequent photon emission by secondary electrons and positrons also can spoil BEs.
Therefore in the simulations described below $x_L / c =  17 \; \mathrm{fs}$ and $\varepsilon
= 2.76 \; \mathrm{GeV}$
($\epsilon = 5400$) are used. The laser pulse amplitude varies to get different values of $\chi$.

In \fref{higherchi} the results of PIC simulations are shown with light blue lines, and the
solution of BEs in GCFA obtained with \textit{Scintillans} are shown with black lines. Figures of the left
column demonstrate
the electron spectrum $f_e(\epsilon)$ whereas the right column is for the photon energy spectrum
$\epsilon f_{ph}(\epsilon)$.
Figures~\ref{higherchi} (a, b), (c, d) and (e, f) correspond to 
$a_0 = 16.5$, $33$ and $66$, respectively. In GCFA this yields the initial quantum parameter $\chi_0 =
0.3$, $0.6$ and $1.2$. With the increase of the magnetic field strength the photon emission time
decreases ($137 t_{rf} = 3.1$, $1.6$ and $0.8 \; \mathrm{fs}$ while $\tau = 6.4 \; \mathrm{fs}$),
the number of
photons emitted per electron increases, and
the electron distribution tends to lower energies.

From the experimental point of view the mean electron energy
\begin{equation}
    \left< \epsilon_e \right> = \int_0^\infty \epsilon f_e(\epsilon) \, d\epsilon
    \times \left[ \int_0^\infty f_e(\epsilon) \, d\epsilon \right]^{-1}
\end{equation}
and the total energy of gamma photons can characterize the interaction~\cite{Cole18, Poder18}.
These quantities are coupled due to energy conservation, and in order to characterize
the resulting electron and photon beams one can use instead, for instance, $\left< \varepsilon
\right>$, the mean electron energy, and $n_\gamma$, the number of
photons with energy higher than $100 \mathrm{~MeV}$, normalized to the number of electrons, which are given in \tref{table} for all the
simulation results depicted in figures~\ref{fokkerplanck} and~\ref{higherchi}. In the table PIC-MC simulation
results are noted as "PIC" and the \textit{Scintillans}'s results as "Sc". The first line
of the table includes the results of PIC simulation with laser pulse diffraction incorporated
(light blue noisy line 1 in \fref{fokkerplanck}(a)).

Note that for the parameters of figures~\ref{higherchi}(a)-(f) almost all assumptions are violated in which
equations~\eref{corr1} and \eref{corr2} are
derived. Namely, the emission spectrum is not determined by the classical synchrotron formula and
by the critical
frequency $\omega_c$, e.g. already at $\chi \approx 0.2$ the emission power is only $1/2$ of that computed
with the
classical formula~\cite{Berestetskii82}. Also, the central limit theorem is not valid because the
number of the emitted
photons is small and the subsequent photon emission events are dependent (the photon energy
is of the order of the electron energy). Therefore, a fairly good coincidence between solutions
of BEs in GCFA and results of PIC simulations with alternating laser field implies that more
reliable basis for the
correspondence~\eref{corr1}-\eref{corr2} can be found.

\begin{table}
    \caption{\label{table} The mean electron energy and the number of photons (per electron) with
    energy $>100 \mathrm{~MeV}$ for the simulation results shown in figures~\ref{fokkerplanck}
    and~\ref{higherchi}.}
    \begin{indented}
    \item[]\begin{tabular}{@{}lllll}
            \br
              figure
            & $\left< \varepsilon_e \right>_{\mathrm{PIC}} \mathrm{[MeV]}$
            & $\left< \varepsilon_e \right>_{\mathrm{Sc}} \mathrm{[MeV]}$
            & $n_{\gamma, \mathrm{PIC}}$
            & $n_{\gamma, \mathrm{Sc}}$ \\
            \mr
            \ref{fokkerplanck}(a), 1 &  391 &  348 & 0.57 & 0.83 \\
            \ref{fokkerplanck}(a), 2 &  674 &  667 & 0.52 & 0.58 \\
            \ref{fokkerplanck}(a), 3 &  845 &  845 & 0.18 & 0.20 \\
            \ref{higherchi}(a-b)     & 2347 & 2335 & 1.02 & 0.96 \\
            \ref{higherchi}(c-d)     & 1813 & 1807 & 2.10 & 1.91 \\
            \ref{higherchi}(e-f)     & 1037 & 1055 & 3.66 & 3.25 \\
    \end{tabular}
    \end{indented}
\end{table}

\section{Conclusion}
\label{conclusion}

In this paper we start from the Fokker--Planck (FP) regime of the photon emission when the number of
the photons emitted per electron is large and the energy of individual photons is small. For this case
two setups are considered: the photon emission by electrons in a constant magnetic field, and the photon
emission by electrons in the head-on collision with a laser pulse. The resulting electron
spectra are the same in both setups if the setup parameters relate
according to~\eref{corr1} and \eref{corr2}. Thus, in FP regime one can compute the electron
spectrum in a global constant field approximation (GCFA) instead using alternating laser field.
Note that GCFA can be justified also in the supercritical
regime $\chi \gg 1$~\cite{Baumann18}.

PIC simulations demonstrate that the established correspondence~\eref{corr1}-\eref{corr2} can be
used far beyond the FP regime, namely, electron spectrum computed in GCFA fits well the
spectrum computed in the laser field for $\chi
\approx 1$ (hence the photon energy is comparable with the electron energy) or for small numbers
of the emitted photons. Moreover, the spectra of the emitted photons in the constant
magnetic field and in the
laser field
coincides fairly well with each other in these cases.

In GCFA the electron and photon spectra can be efficiently computed from 1D Boltzmann's equations,
for what the \textit{Scintillans} open source code is developed~\cite{Scintillans}. The code
conserves exactly the number of electrons and the total particle energy.
Obviously, in 1D Boltzmann's equations neither pulse diffraction nor collective effects in the electron
beam~\cite{Poder18} are taken into account. However, the mentioned effects are generally small and
can be neglected for typical parameters of laser pulse --- electron beam interactions.

Theoretical analysis of radiation reaction in laser fields, e.g. analysis of the stochasticity effect, is
a quite complicated task and often
requires numerical simulations~\cite{Duclous11, Niel18a, Niel18b, Vranic16}. Radiation
reaction in a constant magnetic field looks much more simple than in the alternating field, thus
GCFA opens perspectives of fully theoretical advance of the topic.

In the context of recent experimental study of radiation reaction~\cite{Cole18, Poder18}, GCFA can
be useful as well. \textit{Scintillans} simulations takes time of seconds opposite to PIC
simulations that takes at least minutes, hence a search of optimal
experimental parameters~\cite{Arran19} can be easier in GCFA. Furthermore, in the experiments with
electron spectrum known before and after the collision~\cite{Poder18}, one can find
the laser pulse amplitude and duration from a comparison of the experimental results with
results of \textit{Scintillans} simulations in a large region of parameters. Such comparison
probably will allow to evade experimental uncertainties and to distinguish clearly different
radiation reaction models.

\ack
This research is supported by the Russian Science Foundation through Grant No. 18-72-00121 and
partially by the Russian Foundation for Basic Research (Grant No.  18-32-01061) (PIC development
and simulations, analysis of the diffraction effect) and by the Ministry of Science and Higher
Education funding (No 0035-2014-0006).

\bibliographystyle{iopart-num}
\bibliography{main}

\end{document}